\documentclass[10pt,conference]{IEEEtran}
\usepackage{cite}
\usepackage{amsmath,amssymb,amsfonts}
\usepackage{algorithmic}
\usepackage{textcomp}
\usepackage{xcolor}
\usepackage{booktabs}
\usepackage{graphicx}

\usepackage{multirow}

\def\BibTeX{{\rm B\kern-.05em{\sc i\kern-.025em b}\kern-.08em
    T\kern-.1667em\lower.7ex\hbox{E}\kern-.125emX}}

\usepackage{fancyhdr}
\fancypagestyle{firstpage}{
\fancyhf{}
\fancyhead[C]{To appear at the International Joint Conference on Neural Networks – IJCNN 2026. Maastricht, Netherlands.} 
}

\begin{document}

\newsavebox{\circcontent}
\newlength{\circdim}
\newcommand{\circled}[1]{%
  \sbox{\circcontent}{\strut #1}%
  \setlength{\circdim}{\wd\circcontent}%
  \ifdim\ht\circcontent>\circdim \setlength{\circdim}{\ht\circcontent}\fi
  \ifdim\dp\circcontent>\circdim \setlength{\circdim}{\dp\circcontent}\fi
  \addtolength{\circdim}{0.2em}%
  \begingroup
    \ooalign{%
      \hbox to \circdim{\hss\resizebox*{!}{\circdim}{\ensuremath{\bigcirc}}\hss}\cr
      \hbox to \circdim{\hss\usebox{\circcontent}\hss}\cr
    }%
  \endgroup
}
\newcommand{\rpoint}[1]{\circled{{\fontfamily{pcr}\selectfont\footnotesize #1}}}

\title{
TRAPTI: Time-Resolved Analysis for SRAM Banking and Power Gating Optimization in Embedded Transformer Inference
\vspace{-1em}
}

 \author{
    \IEEEauthorblockN{
        Jan Klhufek\IEEEauthorrefmark{1},
        Alberto Marchisio\IEEEauthorrefmark{2},
        Vojtech Mrazek\IEEEauthorrefmark{1},
        Lukas Sekanina\IEEEauthorrefmark{1},
        Muhammad Shafique\IEEEauthorrefmark{2}
    }
    \IEEEauthorblockA{
        \IEEEauthorrefmark{1}Brno University of Technology, Brno, Czechia\\
        \IEEEauthorrefmark{2}eBRAIN Lab, Division of Engineering, New York University (NYU) Abu Dhabi, Abu Dhabi, UAE\\
        Email: iklhufek@fit.vut.cz, alberto.marchisio@nyu.edu, mrazek@fit.vut.cz, sekanina@fit.vut.cz, muhammad.shafique@nyu.edu
    }
}

\maketitle
\thispagestyle{firstpage}

\begin{abstract}
Transformers achieve state-of-the-art accuracy across language and vision tasks, but their deployment on embedded hardware is hindered by stringent area, latency, and energy constraints. During inference, performance and efficiency are increasingly dominated by the Key--Value (KV) cache, whose memory footprint grows with sequence length, straining on-chip memory utilization. Although existing mechanisms such as Grouped-Query Attention (GQA) reduce KV cache requirements compared to Multi-Head Attention (MHA), effectively exploiting this reduction requires understanding how on-chip memory demand evolves over time. This work presents \textit{TRAPTI}, a two-stage methodology that combines cycle-level inference simulation with time-resolved analysis of on-chip memory occupancy to guide design decisions. In the first stage, the framework obtains memory occupancy traces and memory access statistics from simulation. In the second stage, the framework leverages the traces to explore banked memory organizations and power-gating configurations in an offline optimization flow. We apply this methodology to GPT-2 XL and DeepSeek-R1-Distill-Qwen-1.5B under the same accelerator configuration, enabling a direct comparison of MHA and GQA memory profiles. The analysis shows that DeepSeek-R1-Distill-Qwen-1.5B exhibits a $2.72\times$ reduction in peak on-chip memory utilization in this setting compared to GPT-2 XL, unlocking further opportunities for power-gating optimization.
\end{abstract}


\section{Introduction}
Transformers have become a standard backbone for language and vision models due to their accuracy and parallelizable structure~\cite{Vaswani2017AttentionIsAll}. These Deep Learning (DL) models rely on self-attention mechanisms to compute interactions between input tokens and achieve high accuracy. However, Transformer inference impose substantial computational and memory requirements. Data-center and server-class deployments often satisfy these requirements using Graphics Processing Units (GPUs) coupled with high memory bandwidth and large memory capacity. However, embedded systems face tightly constrained area, latency, and memory budgets, making efficient Transformer inference more challenging.

\begin{figure}[t]
  \centering
  \includegraphics[width=0.49\columnwidth]{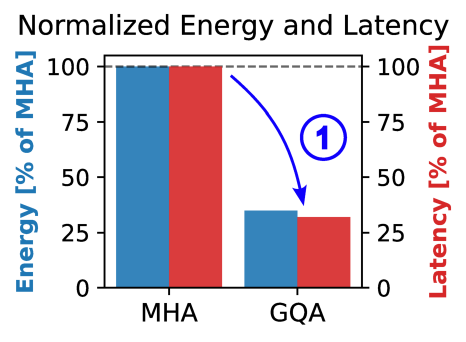}
  \hfill
  \includegraphics[width=0.49\columnwidth]{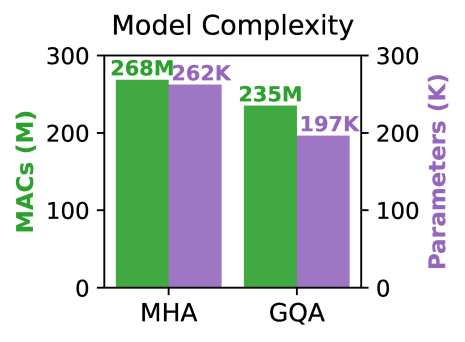}
  \caption{Comparison between Multi-Head Attention (MHA) and Grouped-Query Attention (GQA) in terms of normalized energy and latency at similar parameter count and computational complexity.}
  \label{fig:mha_gqa}
\end{figure}

The inference execution of Transformers consists of repeated self-attention and Feed-Forward Network (FFN) layers, and is dominated by matrix multiplication operations. In decoder architectures that generate output tokens sequentially, the model caches key and value representations from previously generated tokens. This Key--Value (KV) cache grows linearly with the number of generated tokens and can contribute significantly to memory footprint and memory access energy~\cite{Kwon2023KVCachePagedAttention}.

The standard implementation of Multi-Head Attention (MHA) maintains separate key and value tensors for each attention head, which increases both the required memory capacity and the number of memory accesses during decoding. Several attention mechanisms reduce this overhead by decreasing the KV cache footprint, including Multi-Query Attention (MQA)~\cite{Shazeer2019MQA} and Grouped-Query Attention (GQA)~\cite{Ainslie2023GQA}, the latter of which is adopted in recent open-source models targeting efficient inference, such as Qwen~2.5~\cite{Yang2024Qwen25TR} and Llama~3~\cite{Meta2024Llama3}. As illustrated in Fig.~\ref{fig:mha_gqa}, our evaluation shows that GQA yields lower inference energy ($2.89\times$) and latency ($3.14\times$) than MHA at similar parameter count and computational complexity; see pointer~\rpoint{1}.

On embedded inference accelerators, on-chip memory is used to store intermediate activations, weights, and KV cache data, and it can be a major contributor to both dynamic and static energy consumption. 
To avoid performance degradation due to the high off-chip memory traffic, high memory capacity is required.
However, memory utilization during Transformer inference is highly non-uniform over time, creating intervals of time where parts of the memory remain unused. Memory organization techniques such as banking enable partitioning the memory into independently controllable units, creating opportunities for selectively power gating unused banks to reduce leakage energy. \textbf{Effectively exploiting such opportunities requires detailed knowledge of the temporal memory occupancy and access behavior during inference execution.} However, \textit{such temporal information is typically not available to widely used design-space-exploration flows~\cite{Parashar2019Timeloop, Kwon2020MAESTRO, Wu2019Accelergy}}, which often rely on aggregate statistics (e.g., peak capacity or total access counts) rather than execution-aligned occupancy traces. As a result, \textit{prior works have limited ability to systematically identify and evaluate bank-level power-gating opportunities driven by workload dynamics}.

In this work, we present a two-stage framework that combines cycle-level performance simulation with memory-occupancy analysis to guide on-chip memory optimization. 
In Stage~I, we simulate Transformer inference on a given accelerator configuration and extract a time-resolved on-chip memory occupancy profile along with memory access statistics. In Stage~II, we use these extracted traces in an offline exploration to evaluate banked memory organizations and power-gating (PG) policies.
This enables a systematic comparison of energy--latency trade-offs across alternative memory configurations under identical workload behavior. TRAPTI reveals up to 78\% SRAM energy reduction through banking and power-gating, with GQA workloads benefiting 20\% more than MHA due to temporal memory variability.

Our paper introduces the following contributions:
\begin{itemize}
    \item TRAPTI: a two-stage methodology for embedded Transformer inference evaluation that combines cycle-level performance simulation with offline exploration of memory organization and power-gating strategies (Sec.~\ref{sec:methodology}).
    \item Comparative analysis of memory behavior for workloads using MHA versus GQA, highlighting differences in on-chip memory utilization (Sec.~\ref{sec:results}).
    \item A set of banked memory design alternatives evaluated with detailed energy and latency metrics, demonstrating how memory organization choices impact overall energy efficiency (Sec.~\ref{sec:results}).
    \item Integration of the proposed methodology into the TransInferSim tool to support reproducibility and enable future research.
\end{itemize}

Before delving into the technical sections, Sec.~\ref{sec:background} provides an overview of key concepts like attention mechanisms, on-chip memory optimization, accelerator architectures, and inference simulators, to a level of detail necessary to understand the rest of the paper.

\section{Background and Related Work}
\label{sec:background}
\subsection{Attention Mechanisms}
Self-attention is the core operation in Transformers and strongly influences their computational and memory behavior during inference. In standard Transformer architectures, self-attention is implemented using Multi-Head Attention (MHA)~\cite{Vaswani2017AttentionIsAll}. For a sequence of length $N$, forming the $N\times N$ attention matrix leads to $\mathcal{O}(N^2)$ compute and memory complexity. In decoder-only, auto-regressive inference, keys and values from prior tokens are cached to avoid recomputation, resulting in a key--value cache whose size grows linearly with the number of generated tokens and can dominate on-chip storage requirements and traffic~\cite{Kwon2023KVCachePagedAttention}.

\begin{figure}[t]
    \centering
    \includegraphics[width=0.8\columnwidth]{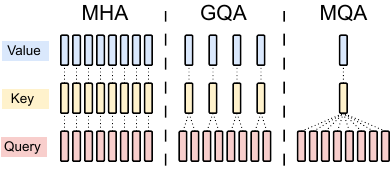}
    \caption{Conceptual comparison of Multi-Head Attention (MHA), Grouped-Query Attention (GQA), and Multi-Query Attention (MQA). MHA maintains independent key and value projections per attention head. GQA shares key and value projections within head groups. MQA shares a single set of key and value projections across all query heads. Figure adapted from~\cite{Ainslie2023GQA}.}
    \label{fig:attention_variants}
\end{figure}

To reduce the memory and bandwidth overhead of KV caching, several attention variants modify how key and value tensors are generated and reused across heads. As summarized in Fig.~\ref{fig:attention_variants}, Multi-Query Attention (MQA) shares a single set of key and value projections across all query heads, reducing KV cache storage and traffic during decoding, but potentially reducing representational flexibility~\cite{Shazeer2019MQA}. Grouped-Query Attention (GQA) introduces an intermediate design point by partitioning query heads into groups that share key and value projections, providing a trade-off between the expressiveness of MHA and the KV cache reduction of MQA~\cite{Ainslie2023GQA}. Since these attention mechanisms directly shape the size and temporal evolution of the KV cache, they also affect how much on-chip memory must remain active throughout decoding.

\subsection{On-Chip Memory Optimization}
\label{subsec:background_memopt}
On-chip Static Random-Access Memory (SRAM) often dominates the energy and area budget of embedded inference accelerators, motivating optimization of both memory capacity and organization. Memory organization parameters, such as banking and port structure, affect achievable access parallelism, contention, and off-chip traffic. Prior work has applied Design Space Exploration (DSE) to Scratchpad Memory (SPM) architectures, including compiler-driven approaches that partition an SPM into banks and selectively disable unused banks to reduce leakage energy~\cite{Kandemir2004BankedSPM}. DESCNet similarly performs application-driven memory DSE and uses utilization profiles to guide multibanked and sectorized SPM organization and low-power control in Capsule Neural Networks accelerators~\cite{Marchisio2021DESCNet}.

Leakage is a major component of SRAM energy in advanced processes and can be reduced by exploiting idle intervals through low-power modes, at the cost of wake-up latency and transition energy. Drowsy caches and cache decay demonstrate that placing unused storage regions into reduced-leakage states can significantly reduce static energy when guided by reuse and idleness behavior~\cite{Flautner2002DrowsyCaches, Kaxiras2001CacheDecay}. For bank-level power gating, energy savings depend on whether idle intervals exceed a break-even duration that amortizes wake-up overhead~\cite{Meng2005ExplLeakLimits,Dropsho2002ManagingStatLeak}. Analytical memory models such as CACTI are widely used to estimate access latency, dynamic energy, leakage power, and area across memory configurations~\cite{Balasubramonian2017CACTI7, Li2011CACTI-P}, and are commonly integrated into architecture-level energy estimation frameworks such as Accelergy~\cite{Wu2019Accelergy}. However, realizing these savings in an accelerator setting requires knowing not only peak capacity, but also when (and for how long) different portions of the memory are idle during execution.

\begin{figure*}[t]
    \centering
    \includegraphics[width=1.0\textwidth]{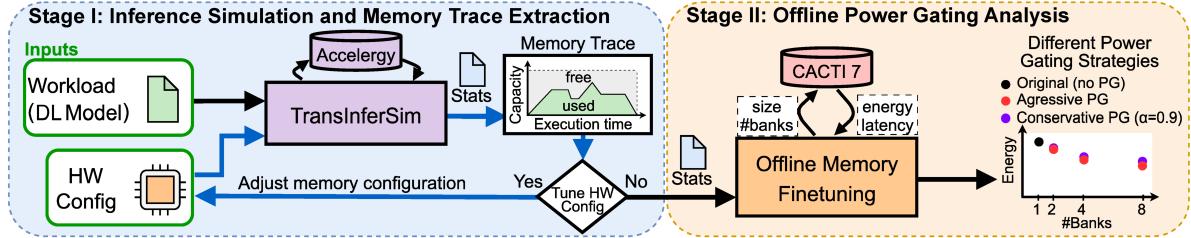}
    \caption{Two-stage workflow for on-chip memory sizing, banking, and power-gating analysis. \textbf{Stage I} (blue) uses TransInferSim to simulate inference for a given workload and accelerator configuration, producing performance statistics and a time-resolved on-chip memory occupancy trace; the blue loop indicates iterative adjustment of the memory configuration until a target SRAM capacity is reached. \textbf{Stage II} (orange) reuses the Stage-I trace and statistics to evaluate banked SRAM organizations and power-gating policies using CACTI~7 characterization, yielding energy trends across bank counts. The power-gating policies include (i) a baseline without power gating, (ii) an \emph{aggressive} policy that gates idle-eligible banks using minimal headroom (\(\alpha\!\approx\!1\)), and (iii) a \emph{conservative} policy that reserves per-bank headroom (\(\alpha\!<\!1\)) and avoids gating across short idle intervals to reduce wake-up overhead.}
    \label{fig:infersim_scheme}
\end{figure*}

\subsection{Accelerator Architectures and Inference Simulators}
Embedded accelerators for Transformer inference typically consist of an array of interconnected Processing Elements (PEs) organized as a Systolic Array (SA). Each PE integrates a Multiply--Accumulate Circuit (MAC) unit and local registers to store intermediate partial results, enabling efficient execution of matrix multiplications. Data movement between the compute array and on-chip memory is managed through a memory hierarchy that may include one or more levels for buffering weights, activations, and intermediate results. Many accelerator designs following this organization have been proposed for Transformer inference~\cite{Qu2022DOTA, Ham2021ELSA, Marchisio2023SwiftTron, Wang2024SOFA}.

Performance evaluation tools for DL accelerators span Register-Transfer Level (RTL) simulation, cycle-level simulation, and analytical or mapping-based estimation, trading off modeling fidelity and evaluation runtime. RTL simulation provides the highest accuracy but is slow and requires a complete hardware description. Cycle-level simulators, such as STONNE~\cite{MunozMartinez2021STONNE}, capture microarchitectural timing and contention with significantly lower cost than RTL. Analytical and mapping-based frameworks, including Timeloop~\cite{Parashar2019Timeloop} and MAESTRO~\cite{Kwon2020MAESTRO}, provide fast latency and energy estimates from dataflow and reuse analysis and are widely used for convolutional and dense-tensor workloads, but rely on aggregate or statistical models of memory behavior. Recent tools target Transformer inference more directly, either at system scale or as compiler-driven flows~\cite{Zhang2024LLMCompass,Mohammad2024CHOSEN}. TransInferSim complements these approaches by performing discrete-event, cycle-level simulation of Transformer inference on custom accelerators, enabling time-resolved analysis of on-chip memory occupancy~\cite{Klhufek2025TransInferSim}.

\noindent
\textbf{Gap and motivation:} While prior work has explored banked memories, low-power modes, and break-even analysis, accelerator evaluation flows often lack a systematic way to translate workload-driven, cycle-accurate execution into time-resolved SRAM demand suitable for bank-level power-gating decisions. In particular, without an occupancy trace aligned to the execution schedule, it is difficult to determine how many banks can be gated at each point in time and whether idle intervals are long enough to outweigh transition overhead. \textit{This paper addresses this gap by extracting an SRAM occupancy trace from cycle-level simulation and reusing it in an offline exploration of banking and power-gating policies.}

\section{TRAPTI Methodology}
\label{sec:methodology}
We propose TRAPTI, a two-stage workflow for on-chip memory sizing, banking, and power-gating analysis in embedded Transformer inference accelerators. Fig.~\ref{fig:infersim_scheme} summarizes the proposed flow. Stage~I uses cycle-level performance simulation to (i) evaluate inference execution for a given Transformer workload and accelerator configuration and (ii) extract a time-resolved on-chip memory occupancy trace. Stage~II performs an offline memory exploration that reuses the trace from Stage~I to evaluate banked SRAM organizations and power-gating configurations. The key idea is to decouple (a) execution timing and data movement, which depend on the workload schedule and accelerator microarchitecture, from (b) SRAM banking and leakage-reduction choices, which can be explored efficiently once time-resolved memory demand is known.

\subsection{Stage I: Simulation and Memory Occupancy Profiling}
\label{sec:methodology_stage1}
\subsubsection{Simulator and input specification}
We use TransInferSim~\cite{Klhufek2025TransInferSim} tool to evaluate inference execution for a given (i) Transformer workload and (ii) accelerator configuration. The workload is provided as a structural description (operation types, tensor dimensions, and dependencies), and the accelerator is specified as one or more compute arrays coupled with an on-chip memory hierarchy (SRAM and local buffers) and an off-chip Dynamic Random Access Memory (DRAM). The main objective of Stage~I is to obtain a cycle-level execution timeline together with a temporal SRAM occupancy trace under the given workload and hardware configuration.

\subsubsection{Execution modeling}
Given the workload description, the simulator constructs an execution plan and performs discrete-event simulation to retrieve the resulting compute and data-movement behavior at cycle-level granularity. This stage captures the interaction between scheduled compute operations and memory accesses, exposing capacity constraints and memory-induced stalls during inference.

\subsubsection{Occupancy trace extraction}
A key artifact produced by the simulation is the time-resolved on-chip SRAM occupancy trace. During simulation, the engine tracks tensors as \emph{needed} (required by future operations) or \emph{obsolete} (no longer required for any future computation). When the memory component becomes full, the simulator selects eviction victims to free space. In our experiments, we configure the memory model to use a Least-Recently Used (LRU) victim selection policy among eligible candidates. Since obsolete tensors are eligible for eviction without impact on correctness, this policy effectively prioritizes evicting obsolete data over needed data when both are present.

If no obsolete data is available, the simulator writes back the needed tensors to an upper memory level (e.g., DRAM) for later refetching. Such write-backs induced by limited memory capacity introduce additional memory traffic and therefore increase latency and energy. Accordingly, we size the SRAM large enough to avoid these capacity-induced write-backs, i.e., to retain the needed tensors throughout execution.

We determine the on-chip memory size by iteratively adjusting its capacity and rerunning simulation until the memory trace reports feasible execution without capacity-induced write-backs. The resulting configuration defines the baseline SRAM size used for the subsequent offline banking and power-gating exploration.

\subsubsection{Stage-I outputs}
Stage~I outputs (i) the time-resolved SRAM occupancy trace and (ii) summary memory access statistics (e.g., read/write counts) under the selected capacity. These outputs are reused directly in Stage~II to evaluate alternative banked SRAM organizations and power-gating policies without modifying the workload execution schedule.

\subsection{Stage II: SRAM Banking and Power-Gating Exploration}
\label{sec:methodology_stage2}

\subsubsection{Banked SRAM candidates and characterization}
We explore alternative banked SRAM organizations while keeping Stage~I execution behavior fixed. For a selected total SRAM capacity $C$ (bytes), we sweep candidate bank counts $B\in\{1,2,4,\ldots\}$, corresponding to an equal-size partitioning of the SRAM into $B$ banks. For each candidate, we use CACTI to obtain bank-level access latency/energy and leakage estimates under the corresponding organization and technology assumptions~\cite{Balasubramonian2017CACTI7,Li2011CACTI-P}.

\subsubsection{Mapping occupancy traces to bank activity}
We translate the SRAM occupancy trace $o(t)$ (bytes) into the minimum number of banks that must remain active at each time $t$. For a total SRAM capacity $C$ and $B$ banks (per-bank capacity $C/B$), we assume occupied data is packed contiguously across banks without fragmentation. To avoid overly optimistic packing, we introduce a per-bank headroom factor $\alpha\in(0,1]$ and treat only $\alpha\cdot(C/B)$ as usable capacity per bank. This accounts for non-ideal placement effects and control/metadata overhead, and provides a simple way to model conservative versus aggressive assumptions. The resulting number of required active banks is
\begin{equation}
B_{\mathrm{act}}(t)=\left\lceil \frac{o(t)}{\alpha\cdot(C/B)}\right\rceil,
\label{eq:Bact}
\end{equation}
where $B_{\mathrm{act}}(t)$ is bounded to the range $0 \le B_{\mathrm{act}}(t) \le B$. The remaining $B-B_{\mathrm{act}}(t)$ banks are considered idle-eligible at time $t$. In this work, we use $\alpha=0.9$ as a conservative guard-band (i.e., reserving 10\% slack per bank), while an aggressive policy can be modeled with $\alpha=1.0$.

\subsubsection{Offline evaluation of banking and power-gating options}
For each $(C,B)$ candidate, we compute the total SRAM energy as the sum of dynamic access energy, leakage energy, and switching overhead:
\begin{equation}
E_{\mathrm{tot}} = E_{\mathrm{dyn}} + E_{\mathrm{leak}} + E_{\mathrm{sw}} .
\label{eq:Etot}
\end{equation}

\textbf{Dynamic energy} is computed from the Stage~I memory-access statistics and per-access energies obtained from CACTI for the corresponding banked organization:
\begin{equation}
E_{\mathrm{dyn}} = N_R \cdot E_R + N_W \cdot E_W ,
\label{eq:Edyn}
\end{equation}
where $N_R$ and $N_W$ denote the total numbers of SRAM reads and writes reported by the simulator, and $E_R$ and $E_W$ are the energy per SRAM read and write from CACTI.

\textbf{Leakage energy} depends on how many banks remain active over time under the selected policy. We use the bank-activity function $B_{\mathrm{act}}(t)$ from Eq.~\eqref{eq:Bact} together with the per-bank leakage power $P_{\mathrm{leak}}^{\mathrm{bank}}$ obtained from CACTI. Using the piecewise-constant segments of the occupancy trace (durations $\Delta t_k$), we estimate:
\begin{equation}
E_{\mathrm{leak}} \approx \sum_{k} P_{\mathrm{leak}}^{\mathrm{bank}} \cdot B_{\mathrm{act}}(k)\cdot \Delta t_k .
\label{eq:Eleak}
\end{equation}

\textbf{Switching overhead} accounts for bank state transitions. Let $N_{\mathrm{sw}}$ be the number of on$\leftrightarrow$off transitions inferred from the bank activity timeline and $E_{\mathrm{sw}}^{\mathrm{bank}}$ the per-transition energy (from CACTI assumptions). We model:
\begin{equation}
E_{\mathrm{sw}} = N_{\mathrm{sw}} \cdot E_{\mathrm{sw}}^{\mathrm{bank}} .
\label{eq:Esw}
\end{equation}

Power gating is applied only when it is beneficial after accounting for switching overhead. For each idle-eligible interval of duration $\Delta t$, a bank is gated only if the expected leakage saved during $\Delta t$ exceeds the wake-up/transition cost and the added wake-up latency is acceptable, following the standard break-even criterion described in Subsection~\ref{subsec:background_memopt}. The outcome is a set of bank-count and gating-policy candidates characterized by their energy and latency implications under the same simulated workload timeline.

\section{Evaluation of TRAPTI}
\label{sec:results}

\subsection{Experimental Setup}
\label{subsec:exp_setup}

\textbf{Workloads.} We evaluate two Transformer workloads with comparable scale: GPT-2 XL (MHA)~\cite{Radford2019GPT2XL} and DeepSeek-R1-Distill-Qwen-1.5B (GQA)~\cite{Guo2025DeepSeekR1}, denoted as \textit{DS-R1D Q-1.5B}. Both workloads are simulated with sequence length $M{=}2048$. Positional-encoding operations (e.g., absolute/rotary position embedding transforms) are omitted from the workload graph because they are element-wise and do not materially affect the SRAM occupancy trends studied in this paper. This simplification is applied consistently across both models. Table~\ref{tab:exp_setup_models} lists the main model hyperparameters: $L$ (layers), $D$ (embedding dimension), $D_{\mathrm{ff}}$ (FFN hidden dimension), $H$ (query heads), $H_{\mathrm{kv}}$ (shared key/value heads for GQA), parameter count $P$ (in Billions), and total MACs (in Trillions). Across all experiments, we uniformly use 8-bit quantized operands.

\begin{table}[t!]
    \centering
    \caption{Model configurations used in the experimental evaluation.}
    \resizebox{1.0\columnwidth}{!}{%
        \begin{tabular}{l|c|c|c|c|c|c|c|c|c|c}
            \toprule
            \textbf{Model} & \textbf{M} & \textbf{L} & \textbf{D} & \textbf{D$_\mathrm{ff}$} &
            \textbf{Attn.} & \textbf{H} & \textbf{H$_\mathrm{kv}$} & \textbf{FFN Type} &
            \textbf{P (B)} & \textbf{MACs (T)} \\
            \midrule
            GPT-2 XL &
            2048 & 48 & 1600 & 6400 &
            MHA & 25 & \textcolor{gray}{--} & FFN &
            1.48 & 3.66 \\
            DS-R1D Q-1.5B &
            2048 & 28 & 1536 & 8960 &
            GQA & 12 & 2 & SwiGLU &
            1.31 & 3.04 \\
            \bottomrule
        \end{tabular}%
    }
    \label{tab:exp_setup_models}
\end{table}

\begin{figure}[t!]
    \centering
    \includegraphics[width=\columnwidth]{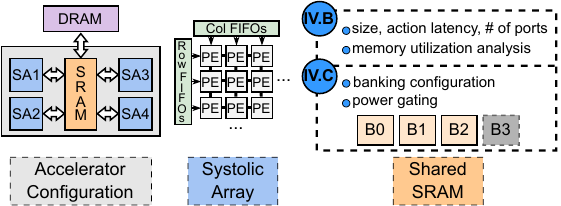}
    \caption{Accelerator design template used in the experimental evaluation. The design comprises four $128\times128$ systolic arrays, each fed by per-array row/column FIFO buffers, and a shared on-chip SRAM connected to off-chip DRAM.}
    \label{fig:hw_exp_setup}
\end{figure}

\textbf{Hardware template.} Fig.~\ref{fig:hw_exp_setup} shows the evaluated accelerator. The compute subsystem consists of four identical SAs, each of size $128\times128$ PEs, operating at 1\,GHz frequency with one 8-bit MAC per cycle per PE. This corresponds to a peak theorethical throughput of $65.5$~TMAC/s. Each SA is fed by a row and a column First-In First-Out (FIFO) buffer stack, each providing 128 lanes with a depth of 256 8-bit elements ($128\times256$ entries). These FIFOs model continuous, pipelined data supply along the tiled inner (common) dimension during matrix multiplication.

\textbf{Memory subsystem.} The baseline memory hierarchy includes a single shared on-chip SRAM with an initial capacity of 128\,MiB, a 512-bit interface with four physical ports, and an access latency of 32~ns, operating at 1\,GHz frequency. The off-chip memory is modeled as a DRAM with 2\,GiB capacity, two physical ports, and 80~ns access latency.

\textbf{TransInferSim settings.} We use TransInferSim with operation sub-tiling set to \texttt{subops}{=}4, which decomposes large matrix multiplications into smaller sub-operations that can be scheduled across the four SAs. This reduces single-array serialization for otherwise wide FFN layers and improves parallel utilization of the compute subsystem.

\textbf{CACTI settings.} We use CACTI with a 45\,nm technology node (consistent with the Accelergy/TransInferSim setup) and the \texttt{itrs-hp} device model for high-performance transistors. All remaining SRAM parameters (e.g., capacity, port count, and interface width) match the TransInferSim memory configuration.

\subsection{Memory Size Optimization}
\label{subsec:mem_size_opt}
We first evaluate the on-chip memory requirements of both workloads under an identical accelerator configuration. The goal of this experiment is to (i) extract time-resolved SRAM occupancy profiles, (ii) identify peak memory requirements, and (iii) relate memory demand to inference time and energy profiles.

\begin{figure*}[h]
  \centering
  \includegraphics[width=\textwidth]{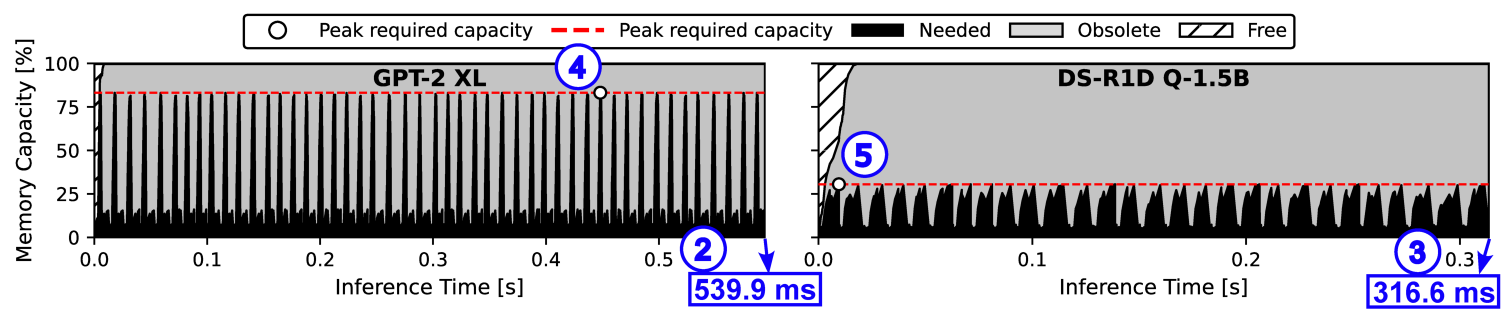}
  \caption{Time-resolved SRAM occupancy for two workloads executed on the same accelerator with a 128\,MiB shared SRAM. \textbf{Left:} GPT-2~XL with MHA (48 layers). \textbf{Right:} DS-R1D~Q-1.5B with GQA (28 layers). DS-R1D~Q-1.5B exhibits lower peak SRAM requirement with more room for optimization.}
  \label{fig:mem_profiles_128}
\end{figure*}

Fig.~\ref{fig:mem_profiles_128} shows the time-resolved SRAM occupancy traces for GPT-2~XL (left) and DS-R1D~Q-1.5B (right), both executed with a 128\,MiB shared SRAM. The corresponding simulated end-to-end inference times are 593.9\,ms for GPT-2~XL (see pointer~\rpoint{2}) and 313.6\,ms for DS-R1D~Q-1.5B (see pointer~\rpoint{3}). The traces decompose the SRAM contents into \emph{needed}, \emph{obsolete}, and \emph{free} regions over time, while highlighting the peak required capacity over the whole simulated run.

GPT-2~XL exhibits substantially higher and more persistent memory occupancy, with a peak requirement of 107.3\,MiB, which is 84\% of the full SRAM capacity; see pointer~\rpoint{4}. This behavior is driven by the larger KV cache footprint of MHA, which causes a significant fraction of the SRAM to remain occupied by needed tensors throughout execution. In contrast, DS-R1D~Q-1.5B shows a markedly lower peak SRAM requirement of 39.1\,MiB (31\% of total SRAM capacity, see pointer~\rpoint{5}), representing a $2.72\times$ reduction in peak utilization compared to GPT-2~XL. The GQA mechanism reduces the KV cache footprint, resulting in a more compact and periodically releasing memory demand profile. This creates extended intervals where a substantial portion of the SRAM remains unused, suggesting more room for further memory size optimization.

\begin{figure}[h]
  \centering
  \includegraphics[width=\columnwidth]{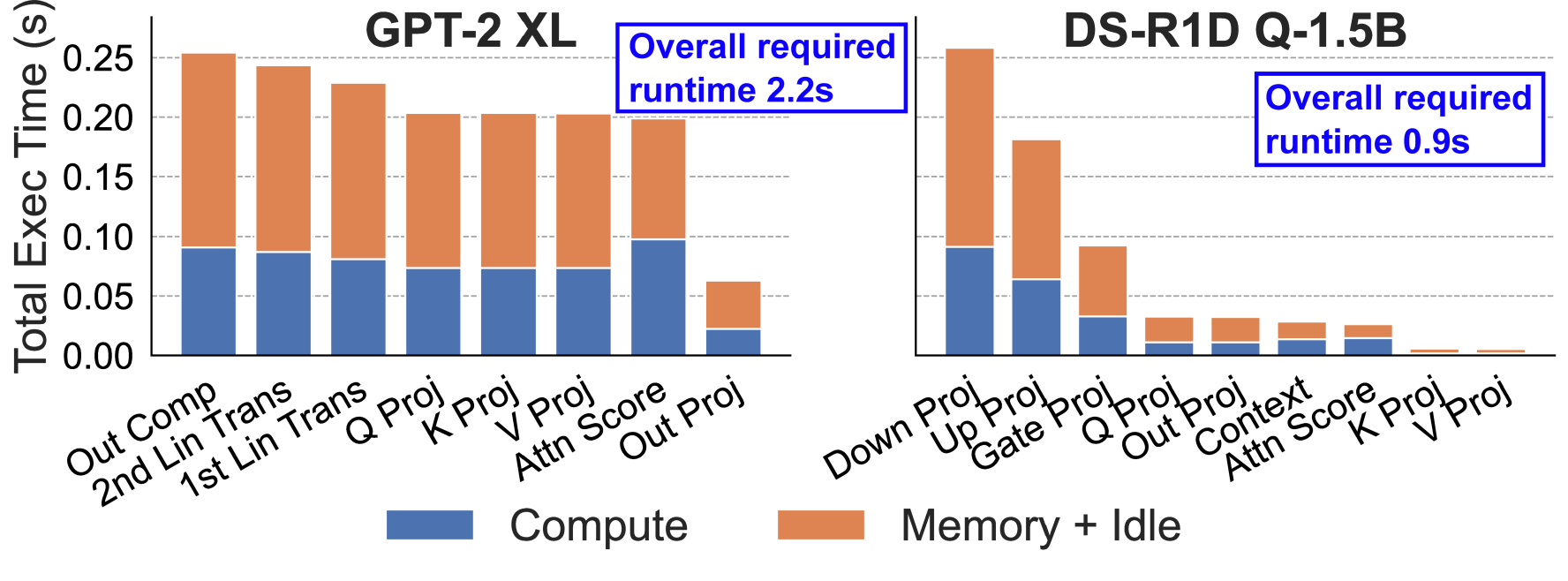}

  \caption{Per-operation latency breakdown for GPT-2~XL (\textbf{left}) and DS-R1D~Q-1.5B (\textbf{right}) under the same accelerator configuration with a 128\,MiB shared SRAM. Each bar separates compute time from memory-access and idle components. GPT-2~XL shows a larger memory/idle fraction across several operations.}
  \label{fig:latency_per_op_type}
\end{figure}

To understand how the different model operations impact execution behavior, Fig.~\ref{fig:latency_per_op_type} reports the aggregated execution times per operation type, separating pure compute time from memory accesses and idle intervals. Note that, since many operations execute in parallel over the available systolic arrays, the total sum remains lower than the total inference time reported in Fig.~\ref{fig:mem_profiles_128}.

For GPT-2~XL, several operations show a large gap between compute-only and memory related latency, indicating frequent stalls and memory-induced delays. This is consistent with the high and sustained SRAM occupancy observed earlier, suggesting that execution is strongly memory bound. In contrast, DS-R1D~Q-1.5B exhibits a much smaller gap between compute and total latency across most operations.

\begin{figure}[h]
  \centering
  \includegraphics[width=\columnwidth]{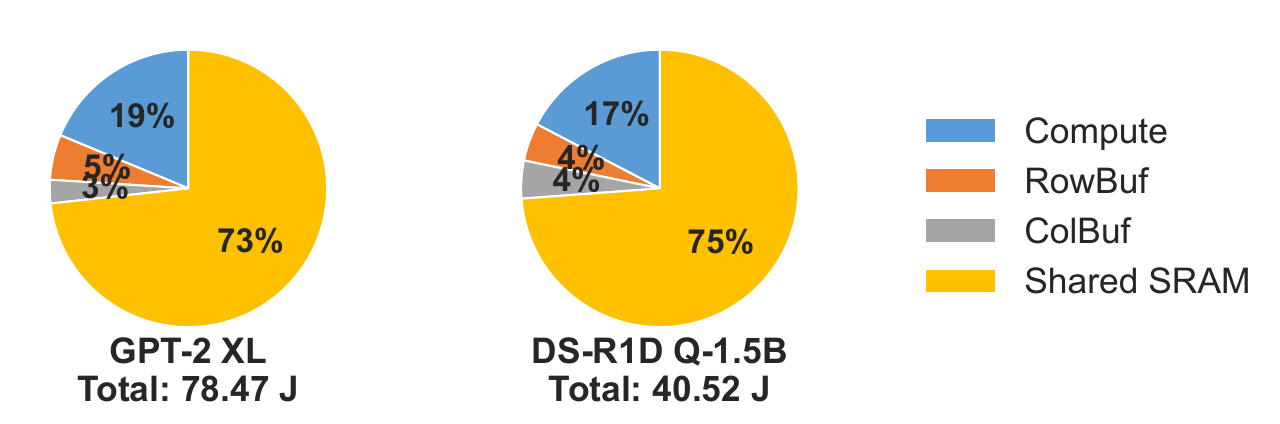}
  \caption{On-chip energy breakdown for GPT-2~XL (\textbf{left}) and DS-R1D~Q-1.5B (\textbf{right}) executed on the same accelerator with a 128\,MiB shared SRAM. DS-R1D~Q-1.5B achieves higher compute utilization and lower total energy, consistent with reduced memory-bound behavior.}
  \label{fig:energy_pies}
\end{figure}

From the perspective of energy consumption, Fig.~\ref{fig:energy_pies} compares the on-chip energy breakdown for both workloads under the same 128\,MiB SRAM configuration. GPT-2~XL consumes 78.47\,J total on-chip energy with 38\% average PE utilization, while DS-R1D~Q-1.5B consumes 40.52\,J with 77\% average PE utilization. The operation breakdown confirms that GPT-2~XL spends a larger fraction of time on memory accesses and idle intervals while the systolic arrays remain under-utilized. DS-R1D Q-1.5B's higher compute utilization translates to more efficient energy use. 
This difference aligns with the observed occupancy and execution time behavior and suggests that GQA reduces memory pressure, allowing execution to operate closer to the accelerator's compute capability.

Based on the memory trace observed for DS-R1D~Q-1.5B, we evaluated it with a reduced shared SRAM capacity of 64\,MiB. Despite halving the SRAM size, the execution latency decreased by only 1.48\,ms compared to the 128\,MiB configuration. This minimal latency impact can be explained by the occupancy trace as the peak memory requirement of DS-R1D~Q-1.5B remains well below 64\,MiB, with no capacity-induced write-backs. The small latency difference is primarily due to the modeled access latency of the smaller SRAM (22~ns for 64\,MiB), rather than changes in execution behavior or memory traffic.

Since the latency impact is negligible and the occupancy trace provides the peak required capacity (48\,MiB for DS-R1D~Q-1.5B, 112\,MiB for GPT-2~XL), we leverage this information for the subsequent Stage~II to evaluate a range of SRAM sizes (with capacities above the minimum requirement), sweeping upward in 16\,MiB increments up to 128\,MiB to quantify how added headroom shifts banking and power-gating behavior.

\subsection{Power Gating Optimization}
\label{subsec:pg_opt}

We next apply Stage~II of our methodology to translate the time-resolved SRAM occupancy traces into energy and area trade-offs across banking and power-gating configurations. We first illustrate how the occupancy trace is mapped to bank-level activity, and then quantify the resulting energy--area trends across SRAM capacities and bank counts.

\begin{figure}[t]
  \centering
  \includegraphics[width=\columnwidth]{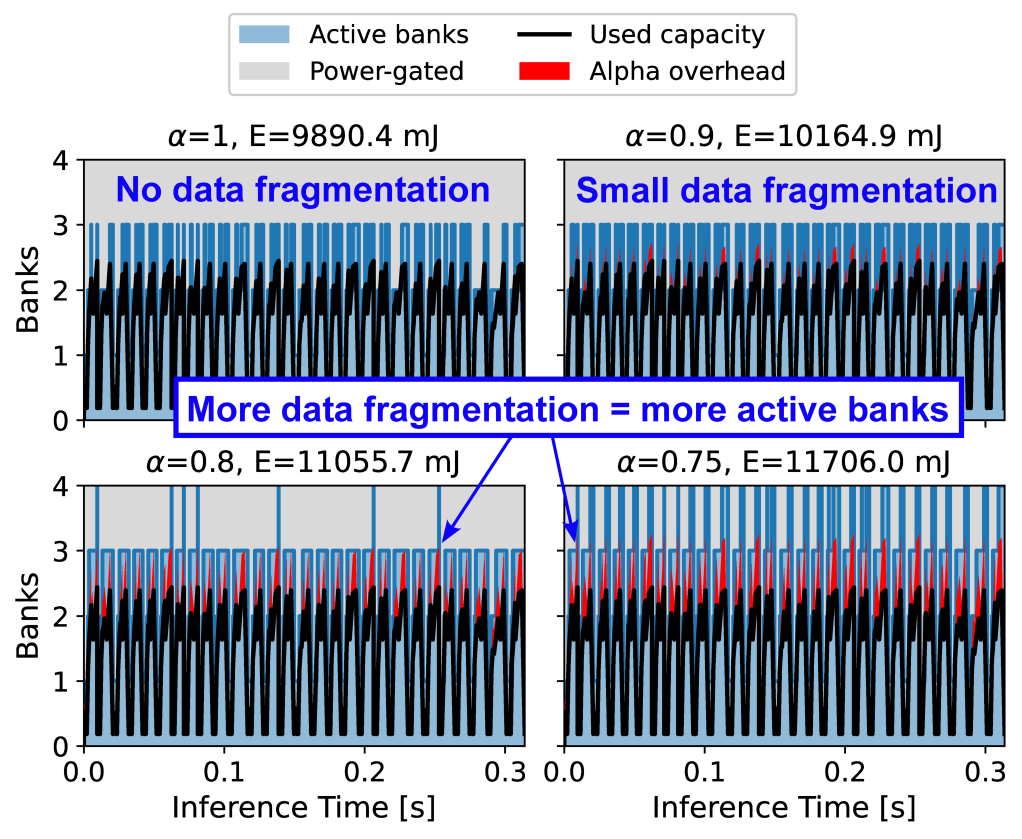}
  \caption{Inferred bank activity timeline for DS-R1D~Q-1.5B at 64\,MiB with $B{=}4$ under different values of $\alpha$. The \textbf{black} curve shows total SRAM occupancy over time. The \textbf{red} region represents the effective data-placement overhead induced by $\alpha$, which reduces the usable per-bank capacity and models conservative assumptions about non-ideal data packing. \textbf{Blue} segments indicate banks that must remain active. Smaller $\alpha$ values increase the modeled overhead, resulting in more active banks and fewer gate-eligible intervals.}
  \label{fig:pg_timeline_facets}
\end{figure}

Fig.~\ref{fig:pg_timeline_facets} shows the inferred bank activity over time for DS-R1D~Q-1.5B at 64\,MiB with $B{=}4$ under different values of the headroom factor $\alpha$. The plot illustrates how varying $\alpha$ changes the number of banks that must remain active over time while keeping the workload execution and total SRAM capacity fixed.

The factor $\alpha$ controls how conservatively the total SRAM occupancy is mapped to bank-level activity. At $\alpha{=}1.0$, ideal data packing maximizes idle banks during low-occupancy intervals.
Lower $\alpha$ values introduce a conservative margin that models uncertainty in data placement and implementation overheads. As $\alpha$ decreases, more banks are required to remain active for the same occupancy trace, reducing the duration and frequency of gate-eligible intervals. Bank activity tracks the occupancy trace across all $\alpha$ values as memory demand fluctuates.

Based on this behavior, we fix $\alpha{=}0.9$ for the remainder of the evaluation as a conservative yet practical setting that avoids optimistic packing assumptions while still preserving meaningful power-gating opportunities. Using this fixed $\alpha$, we sweep SRAM capacities and bank counts $B\in\{1,2,4,8,16,32\}$ to quantify energy and area trade-offs.

\begin{table*}[ht]
\centering
\caption{Energy/area trade-offs for SRAM banking candidates at $\alpha$=0.9. Energy (E) in mJ, area (A) in mm$^2$, varying the number of banks~(B); $\Delta$ values are relative to B=1 for each memory capacity (C). For DS-R1D~Q-1.5B, banking reduces SRAM energy across all capacities, with the strongest reductions at $B\in\{8,16\}$ (\textbf{bold}), while further banking yields diminishing energy returns and area overhead. GPT-2~XL shows similar trends but is evaluated only at larger capacities due to its higher SRAM requirement.}
\label{tab:sram_banking_energy_area}
\resizebox{\textwidth}{!}{%
\begin{tabular}{c|cc|cccc|cccc|cccc|cccc|cccc}\toprule
\multicolumn{23}{c}{\textbf{DeepSeek-R1-Distill-Qwen-1.5B}} \\
\midrule
\multirow{2}{*}{\centering C [MiB]} & \multicolumn{2}{c|}{B=1} & \multicolumn{4}{c|}{B=2} & \multicolumn{4}{c|}{B=4} & \multicolumn{4}{c|}{B=8} & \multicolumn{4}{c|}{B=16} & \multicolumn{4}{c}{B=32} \\
 & E & A & E & A & $\Delta$E & $\Delta$A & E & A & $\Delta$E & $\Delta$A & E & A & $\Delta$E & $\Delta$A & E & A & $\Delta$E & $\Delta$A & E & A & $\Delta$E & $\Delta$A \\
\midrule
48 & 12767.6 & 854.50   & 10733.1 & 876.01 & -15.9 & +2.5 & 9323.6 & 911.27 & -27.0 & +6.6 & \textbf{8883.9} & 927.88 & -30.4 & +8.6 & 9181.2 & 934.40 & -28.1 & +9.4 & 9787.5 & 954.81 & -23.3 & +11.7 \\
64 & 15912.9 & 1126.74   & 11932.1 & 1187.50 & -25.0 & +5.4 & 10164.9 & 1218.42 & -36.1 & +8.1 & 9759.2 & 1235.18 & -38.7 & +9.6 & \textbf{9621.7} & 1287.32 & -39.5 & +14.2 & 10248.6 & 1321.18 & -35.6 & +17.3 \\
80 & 20085.8 & 1432.50   & 13385.8 & 1457.78 & -33.4 & +1.8 & 11667.3 & 1490.73 & -41.9 & +4.1 & 10325.6 & 1618.55 & -48.6 & +13.0 & \textbf{10181.8} & 1645.72 & -49.3 & +14.9 & 10606.6 & 1729.22 & -47.2 & +20.7 \\
96 & 23672.2 & 1696.02   & 13839.5 & 1728.07 & -41.5 & +1.9 & 12387.5 & 1763.04 & -47.7 & +4.0 & 10963.2 & 1909.79 & -53.7 & +12.6 & \textbf{10727.5} & 1944.08 & -54.7 & +14.6 & 11342.2 & 2091.05 & -52.1 & +23.3 \\
112 & 27258.8 & 1959.54   & 15898.5 & 2081.32 & -41.7 & +6.2 & 13959.7 & 2035.35 & -48.8 & +3.9 & 12053.8 & 2201.03 & -55.8 & +12.3 & \textbf{11461.7} & 2322.45 & -58.0 & +18.5 & 11725.3 & 2398.99 & -57.0 & +22.4 \\
128 & 29904.3 & 2196.94   & 17749.8 & 2280.85 & -40.6 & +3.8 & 13866.3 & 2335.46 & -53.6 & +6.3 & 12083.3 & 2357.82 & -59.6 & +7.3 & \textbf{11584.9} & 2425.46 & -61.3 & +10.4 & 11946.7 & 2556.61 & -60.1 & +16.4 \\
\midrule\midrule
\multicolumn{23}{c}{\textbf{GPT-2 XL}} \\
\midrule
\multirow{2}{*}{\centering C [MiB]} & \multicolumn{2}{c|}{B=1} & \multicolumn{4}{c|}{B=2} & \multicolumn{4}{c|}{B=4} & \multicolumn{4}{c|}{B=8} & \multicolumn{4}{c|}{B=16} & \multicolumn{4}{c}{B=32} \\
 & E & A  & E & A & $\Delta$E & $\Delta$A & E & A & $\Delta$E & $\Delta$A & E & A & $\Delta$E & $\Delta$A & E & A & $\Delta$E & $\Delta$A & E & A & $\Delta$E & $\Delta$A \\
\midrule
112 & 52409.9 & 1959.54   & 35176.7 & 2035.35 & -32.9 & +3.9 & 27358.0 &  2081.32 & -47.8 & +6.2 & 24829.4 & 2201.03 & -52.6 & +12.3 & \textbf{24713.1} & 2322.45 & -52.8 & +18.5 & 25892.2 & 2398.99 & -50.6 & +22.4 \\
128 & 57480.6 & 2196.94   & 38996.0 & 2280.85 & -32.2 & +3.8 & 30022.8 & 2335.46 & -47.8 & +6.3 & 26591.3 & 2357.82 & -53.7 & +7.3 & \textbf{25394.6} & 2425.46 & -55.8 & +10.4 & 26296.8 & 2556.61 & -54.3 & +16.4 \\
\bottomrule
\end{tabular}%
}
\end{table*}

Table~\ref{tab:sram_banking_energy_area} reports CACTI SRAM energy and area estimates for banking candidates evaluated at $\alpha{=}0.9$, with $\Delta$ values computed relative to the unbanked case ($B{=}1$) for each capacity. For DS-R1D~Q-1.5B, increasing the bank count consistently reduces total SRAM energy across all evaluated capacities. The largest energy reductions are observed for $B\in\{8,16\}$. For $B > 16$, additional banking yields diminishing energy returns while continuing to increase area.

GPT-2~XL shows similar trends, but the analysis is restricted to larger capacities (112--128\,MiB) due to its higher peak SRAM requirement. As a result, GPT-2~XL exhibits a narrower design space with fewer opportunities to exploit idle intervals, leading to smaller relative gains from banking compared to DS-R1D~Q-1.5B.

\begin{figure}[h]
  \centering
  \includegraphics[width=\columnwidth]{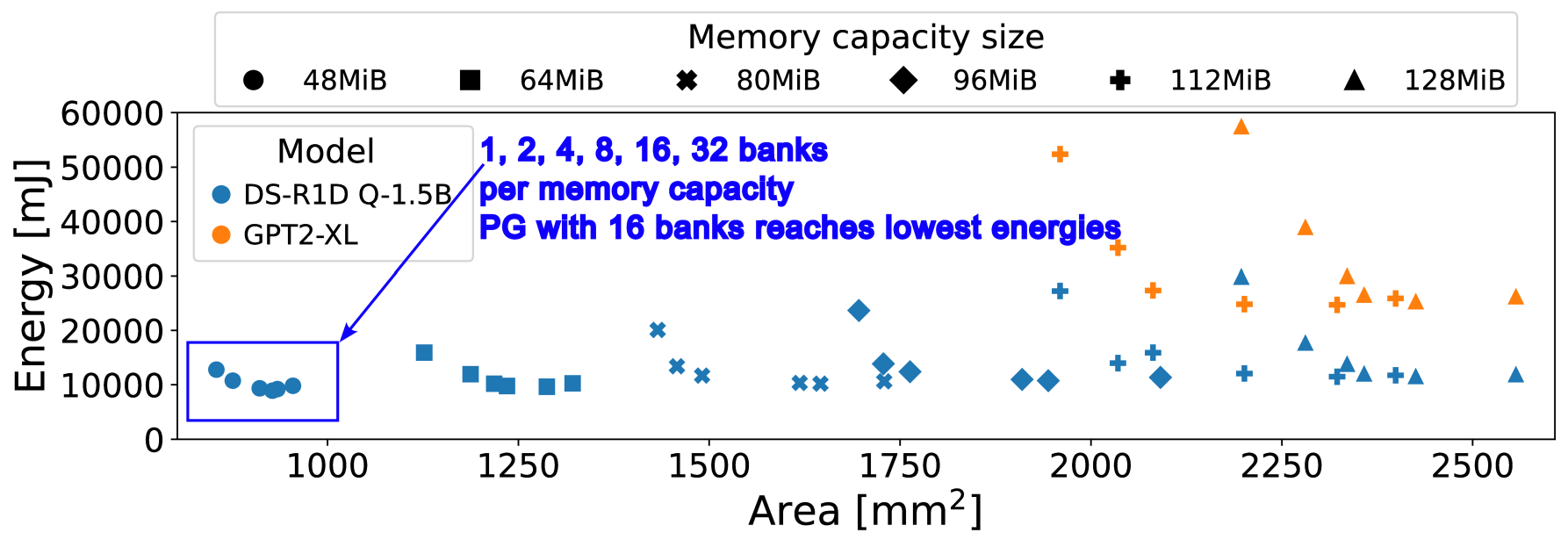}
  \caption{Energy--area trade-off for banked SRAM configurations evaluated at $\alpha{=}0.9$ for both GPT-2~XL (\textbf{orange}) and DS-R1D~Q-1.5B (\textbf{blue}) over various possible memory capacities. Each point corresponds to a different $(C,B)$ candidate.}
  \label{fig:area_energy_scatter}
\end{figure}

Fig.~\ref{fig:area_energy_scatter} summarizes these results by plotting all evaluated $(C,B)$ configurations in terms of SRAM energy and area. The plot highlights the overall trade-off where increasing the number of banks reduces energy at the cost of higher area. The DS-R1D~Q-1.5B workload consistently achieves lower SRAM energy due to its reduced and more variable memory demand. GPT-2~XL's sustained high occupancy limits banking benefits. We also observed that the switching overhead had a negligible impact on overall energy consumption.

These results demonstrate that time-resolved occupancy traces enable systematic SRAM banking evaluation. GQA workloads benefit more from banking due to extended idle intervals, while MHA's sustained occupancy offers limited leakage-reduction opportunities.

\subsection{Multi-level Hierarchy}
\begin{figure}[h]
    \centering
    \includegraphics[width=0.75\columnwidth]{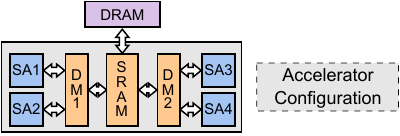}
    \caption{Multi-level on-chip memory hierarchy accelerator setup. In addition to the shared SRAM, two dedicated memories are attached to separate pairs of systolic arrays and to the shared SRAM, which fetches data from DRAM and additionally serves as a backup storage.}
    \label{fig:hw_multilevel}
\end{figure}

To demonstrate the generalization of our approach beyond single-SRAM designs, we evaluate our methodology on a candidate accelerator with multi-level memory hierarchy.
Two Dedicated Memories (DM1 and DM2) are connected to two of the SAs, alongside the shared SRAM (see Fig.~\ref{fig:hw_multilevel}). All three on-chip memories are conservatively sized to 64\,MiB, while the compute subsystem and architectural parameters remain identical to the baseline configuration used in previous experiments.

We evaluate this multi-level memory hierarchy only for DS-R1D~Q-1.5B. The time-resolved occupancy traces indicate peak requirements of 34.1\,MiB for the shared SRAM, and 35.5\,MiB and 37.7\,MiB for DM1 and DM2, respectively, which are slightly below the peak requirement observed in the initial single-memory setup.

Table~\ref{tab:sram_banking_energy_area_memtypes} reports the CACTI-based SRAM energy/area estimates for banking the shared SRAM, DM1, and DM2 independently under the multi-level hierarchy at $\alpha{=}0.9$. Across all three memories, increasing the bank count reduces the estimated SRAM energy, and the relative reduction is larger than that in the single shared-SRAM case for comparable capacities. This occurs because occupancy is distributed across multiple on-chip memories, creating longer and more frequent idle-eligible intervals within each individual memory.

We observed that introducing dedicated memories also creates additional data-movement paths between compute units. In the non-optimized scheduling flow used here, data produced and stored in one dedicated memory may need to traverse the interconnect to another memory before being consumed by a different systolic array. This increased data hopping and coordination overhead leads to higher end-to-end latency (550\,ms), lower average compute utilization (57\%), and higher total on-chip energy consumption (73.4\,J) compared to the single-level configuration. Our trace-based methodology applies to arbitrary memory hierarchies, including multi-level designs, but realizing their full potential requires optimized data placement and scheduling, which are beyond the scope of this work.

\begin{table}[t]
\centering
\caption{Energy and area estimates for banked on-chip memories in the multi-level hierarchy at $\alpha{=}0.9$, evaluated independently for the shared SRAM and both dedicated memories for DS-R1D~Q-1.5B. Energy is reported in mJ and area in mm$^2$; $\Delta$ values are relative to the unbanked configuration ($B{=}1$) for each memory type and capacity.}
\label{tab:sram_banking_energy_area_memtypes}
\Huge
\resizebox{\columnwidth}{!}{%
\begin{tabular}{c|cc|cccc|cccc|cccc}\toprule
\multirow{2}{*}{C [MiB]} & \multicolumn{2}{c|}{B=1} & \multicolumn{4}{c|}{B=4} & \multicolumn{4}{c|}{B=8} & \multicolumn{4}{c}{B=16} \\
 & E & A & E & A & $\Delta$E & $\Delta$A & E & A & $\Delta$E & $\Delta$A & E & A & $\Delta$E & $\Delta$A \\
\midrule
\multicolumn{15}{c}{\textbf{Shared SRAM}} \\
\midrule
48 & 15702.4 & 854.50 & 5885.6 & 911.27 & -62.5 & +6.6 & 4620.0 & 927.88 & -70.6 & +8.6 & \textbf{4216.8} & 934.40 & -73.1 & +9.4 \\
64 & 20171.2 & 1126.74 & 6602.4 & 1218.42 & -67.3 & +8.1 & 5096.0 & 1235.18 & -74.7 & +9.6 & \textbf{4480.0} & 1287.32 & -77.8 & +14.2 \\
\midrule\midrule
\multicolumn{15}{c}{\textbf{Dedicated Memory 1}} \\
\midrule
48 & 18233.6 & 854.50 & 7448.0 & 911.27 & -59.2 & +6.6 & \textbf{6059.2} & 927.88 & -66.8 & +8.6 & 6171.2 & 934.40 & -66.2 & +9.4 \\
64 & 23402.4 & 1126.74 & 8713.6 & 1218.42 & -62.8 & +8.1 & 6708.8 & 1235.18 & -71.3 & +9.6 & \textbf{6451.2} & 1287.32 & -72.4 & +14.2 \\
\midrule\midrule
\multicolumn{15}{c}{\textbf{Dedicated Memory 2}} \\
\midrule
48 & 18636.3 & 854.50 & 8460.4 & 911.27 & -54.6 & +6.6 & \textbf{7068.3} & 927.88 & -62.1 & +8.6 & 7222.9 & 934.40 & -61.2 & +9.4 \\
64 & 23904.7 & 1126.74 & 9674.4 & 1218.42 & -59.5 & +8.1 & 7874.5 & 1235.18 & -67.1 & +9.6 & \textbf{7208.9} & 1287.32 & -69.8 & +14.2 \\
\bottomrule
\end{tabular}%
}
\end{table}

\section{Conclusion}
We presented TRAPTI, a two-stage workflow for analyzing on-chip SRAM sizing, banking, and power gating for embedded Transformer inference using time-resolved memory occupancy traces from simulation. Cycle-level simulation revealed that, under the same accelerator configuration, the GQA-based DeepSeek-R1-Distill-Qwen-1.5B workload requires substantially less peak on-chip SRAM than GPT-2~XL, with a $2.72\times$ lower peak utilization, and achieves a $1.89\times$ faster simulated end-to-end inference time. Using the extracted occupancy traces for offline exploration, we evaluated banked SRAM organizations and power-gating policies and quantified the resulting energy--area trade-offs across feasible memory capacities and bank counts, revealing up to 78\% energy reduction at $\alpha{=}0.9$. Workloads with lower and more variable memory demand (GQA) provide more gate-eligible intervals and therefore benefit more from banking and power gating, while memory-intensive workloads (MHA) offer a narrower design space. Future work will extend the analysis to deeper multi-level on-chip hierarchies and incorporate more detailed transition overhead models and policy sensitivity studies.

\section*{Acknowledgment}
This work was supported in part by the Czech Science Foundation project 25-15490S and the NYUAD Center for Cyber Security (CCS), funded by Tamkeen under the NYUAD Research Institute Award G1104.

\bibliographystyle{ieeetr}
\bibliography{main.bib}

\end{document}